\documentclass[twocolumn,superscriptaddress,prb]{revtex4}
\usepackage{amsmath}
\usepackage{graphicx}
\usepackage{xcolor} 
\usepackage[normalem]{ulem} 

\newcommand{\stkout}[1]{\ifmmode\text{\sout{\ensuremath{#1}}}\else\sout{#1}\fi}
\newcommand{\md}{\mathrm d}

\begin{document}

\title{Optimal control with a strong harmonic trap}

\author{Steven Blaber}
\email{sblaber@sfu.ca}
\affiliation{Dept.~of Physics, Simon Fraser University, Burnaby, British Columbia V5A 1S6, Canada}
\author{David A.\ Sivak}
\email{dsivak@sfu.ca}
\affiliation{Dept.~of Physics, Simon Fraser University, Burnaby, British Columbia V5A 1S6, Canada}

\begin{abstract}
Quadratic trapping potentials are widely used to experimentally probe biopolymers and molecular machines and drive transitions in steered molecular-dynamics simulations. Approximating energy landscapes as locally quadratic, we design multidimensional trapping protocols that minimize dissipation. The designed protocols are easily solvable and applicable to a wide range of systems. The approximation does not rely on either fast or slow limits and is valid for any duration provided the trapping potential is sufficiently strong. We demonstrate the utility of the designed protocols with a simple model of a periodically driven rotary motor. Our results elucidate principles of effective single-molecule manipulation and efficient nonequilibrium free-energy estimation.
\end{abstract}

\maketitle

\emph{Introduction.}---The molecular machine ATP synthase is a remarkable rotary motor that is driven by a proton gradient to synthesize ATP. Single-molecule biophysical experiments have isolated the F$_1$ component of the machine in order to drive it with chemical gradients and applied torques, finding nearly perfect energetic efficiency~\cite{kinosita2000,Toyabe2011}. Magnetic and optical trapping potentials such as those used to study ATP synthase~\cite{Toyabe2011,toyabe2012,kawaguchi2014} or carry out DNA folding/unfolding experiments~\cite{liphardt2002,collin2005,bustamante2000,bustamante2003,woodside2006,neupane2017} are well approximated as quadratic. By varying the focus and intensity of the trapping potential, the center and stiffness can be precisely controlled. In this letter, we design energetically efficient driving protocols for strong quadratic trapping potentials on arbitrary energy landscapes. We describe general design principles that can be applied to biophysical experiments and molecular simulations.

We employ the theoretical framework of stochastic thermodynamics, which describes the energy flows in small-scale fluctuating systems~\cite{Jarzynski2011,Seifert2012}. Much like its classical thermodynamic roots~\cite{Carnot1960}, stochastic thermodynamics seeks to describe the design principles of energetically efficient machines, but now at the micro- and nano-scale. Unlike thermodynamics at the macro-scale, small-scale systems are constantly bombarded by relatively large fluctuations, typically operate on timescales comparable to their natural relaxation times, and are therefore not well described by the familiar quasistatic processes often used to describe efficient machines~\cite{Schroeder}. 

Despite the added complexity, considerable strides have been made towards the general description of efficient stochastic-thermodynamic systems~\cite{Brown2019}. These systems can be driven by two types of \emph{protocols}: \emph{constrained-final-distribution} (CFD) protocols and \emph{constrained-final-control-parameter} (CFCP) protocols. CFD protocols assume complete control over the dynamics and drive the system to a specified final probability distribution, guaranteeing that at any driving speed the system will reach its destination. This can be used to model rotary motors like ATP synthase by setting the final state to be identical to the initial but shifted by one period, resulting in one cycle of free energy transduction in a specified duration. 

For one-dimensional overdamped dynamics, optimal-transport theory can be directly applied to determine the minimum-dissipation protocol that drives a system between specified initial and final distributions~\cite{Aurell2011} and to place fundamental bounds on the cost of information erasure~\cite{Proesmans2020}. For multidimensional overdamped dynamics, the entropy production is lower bounded by the Wasserstein distance between the initial and final distributions~\cite{dechant2022,miangolarra2022}, and exact minimum-dissipation protocols are known for quadratic trapping potentials in isolation~\cite{Schmiedl2007,abiuso2022}. 

CFCP protocols drive a finite set of control parameters to specified final control-parameter values. For such protocols, the system does not necessarily keep up with rapid changes in control parameters, and for fast driving the system state remains largely unchanged~\cite{Blaber2021}. For nonequilibrium free-energy estimation, the free-energy change is estimated from work measurements between control-parameter endpoints, so CFCP protocols are the natural choice. The minimum-dissipation protocol is described by a geodesic of a Riemannian friction-tensor metric when the protocol is sufficiently slow~\cite{OptimalPaths}, by a short-time efficient protocol (STEP) when the protocol is sufficiently fast~\cite{Blaber2021}, and by linear-response theory when the perturbations are sufficiently weak~\cite{bonanca2018,kamizaki2022}. These methods have been used to explore a diverse set of model systems~\cite{Sivak2016,Blaber2020,deffner2020,zulkowski2012,bonancca2014,zulkowski2015,zulkowski2015Quantum,Large2019,Lucero2019,Rotskoff2015,Rotskoff2017,louwerse2022,frim2022,frim2021}, including DNA pulling experiments~\cite{Tafoya2019}, and have been applied to improve free-energy estimation~\cite{Blaber2020Skewed}. 

\begin{figure*}
	\includegraphics[width=\linewidth]{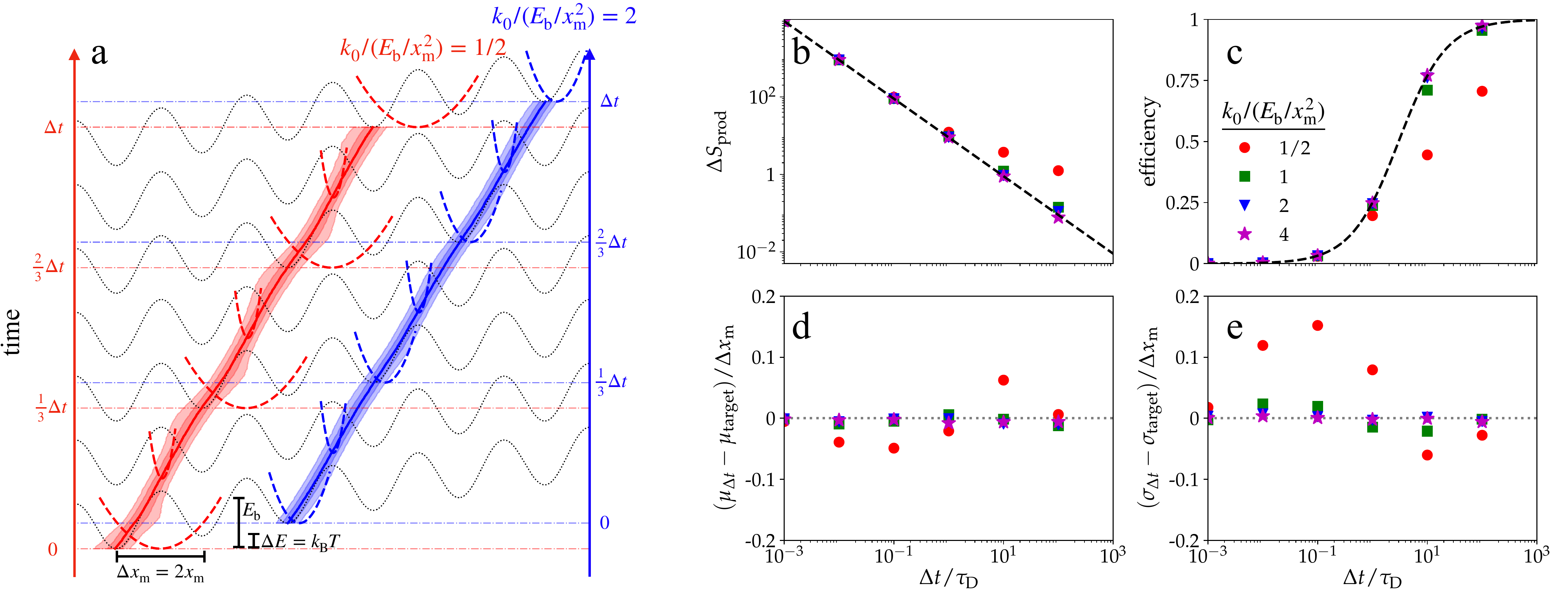}
	\caption{Performance of model rotary motor. (a) Time-dependent protocol for a weak initial stiffness $k_0/(E_{\rm b}/x_{\rm m}^2) = 1/2$ (red, left) and intermediate initial stiffness $k_0/(E_{\rm b}/x_{\rm m}^2) = 2$ (blue, right) depicting the static potential (dotted/gray), trap potential (dashed), median position (solid), and $9\%$, $25\%$, $75\%$, and $91\%$ quantiles (shaded). The two protocols are offset vertically and horizontally for clarity. (b) Entropy production, (c) efficiency, (d) deviation of the final mean position from the target, and (e) deviation of the final standard deviation from the target. Different colors/markers represent different initial stiffnesses $k_{0}$. Black dashed curves in (b) and (c): strong-trap approximations \eqref{entropy production at constant variance} and \eqref{efficiency}, respectively. The energy offset is $\Delta E = k_{\rm B}T$, the barrier height is $E_{\rm b} = 4k_{\rm B}T$, and in (a) the protocol duration is $\Delta t = \tau_{\rm D}$ for diffusion time $\tau_{\rm D} \equiv \Delta x_{\rm m}^2/D$ between adjacent wells. Error bars representing bootstrap-resampled 95\% confidence intervals are smaller than the markers.}
	\label{Protocol}
\end{figure*}

We describe minimum-dissipation protocols for multidimensional overdamped dynamics driven over arbitrary energy landscapes by quadratic trapping potentials. Approximating the static energy landscape as locally quadratic, we obtain minimum-dissipation protocols that are valid for strong trapping potentials at any driving speed. For equal initial and final covariance, the minimum-dissipation CFD protocols are given by explicit  equations for the trap center~\eqref{optimal center} and stiffness~\eqref{optimal stiffness} that linearly drive the mean between the specified endpoints while maintaining constant covariance. We perform a second optimization~\eqref{Optimal Work} to achieve CFCP minimum-dissipation protocols. Since the designed CFD protocols can be solved analytically and calculating CFCP protocols only requires performing a minimization over the final mean and covariance (which in some cases is analytic~\eqref{Optimal Final Mean}), these designed protocols are considerably simpler to determine compared to previous methods~\cite{OptimalPaths,Blaber2021,Aurell2011}. We illustrate our results with a simple model of a rotary motor (Fig.~\ref{Protocol}). By tightening the trap as it crosses energy barriers, the designed protocol achieves minimal entropy production~\eqref{entropy production at constant variance} and maximum efficiency~\eqref{efficiency}, provided the trap is sufficiently strong to confine the system within a single well.

\emph{Minimum-dissipation quadratic control.}---Consider the overdamped motion of a system with diffusion coefficient $D$ driven by a time-dependent potential $V_{\rm tot}(\boldsymbol{r},t)$, satisfying the Fokker-Planck equation
\begin{align}
\frac{\partial p(\boldsymbol{r},t)}{\partial t} = -\nabla\cdot\left[\boldsymbol{v}(\boldsymbol{r},t)p(\boldsymbol{r},t)\right] \ ,
\label{Fokker Planck}
\end{align}
where
\begin{align}
\boldsymbol{v}(\boldsymbol{r},t) \equiv -D\nabla\left[\beta V_{\rm tot}(\boldsymbol{r},t) + \ln p(\boldsymbol{r},t)\right] \ ,
\label{velocity}
\end{align}
is the mean local velocity~\cite{nakazato2021} and $p(\boldsymbol{r},t)$ the system's probability distribution over position vector $\boldsymbol{r}$ at time $t$. The system is in contact with a heat bath at temperature $T$, with $\beta \equiv (k_{\rm B}T)^{-1}$ for Boltzmann's constant $k_{\rm B}$.

The total potential $V_{\rm tot} = V_{\rm land} + V_{\rm trap}$ is separated into a time-independent component $V_{\rm land}$ (the underlying energy landscape) and a quadratic trapping potential
\begin{align}
V_{\rm trap}[\boldsymbol{r},\boldsymbol{\lambda}_{t},K_{t}]=\frac{1}{2}\left[\boldsymbol{r}-\boldsymbol{\lambda}_{t}\right]^{\top} K_{t}\left[\boldsymbol{r}-\boldsymbol{\lambda}_{t}\right] \ .
\end{align}
$K$ is the symmetric stiffness matrix, superscript $\top$ is the vector transpose, and subscript $t$ denotes a variable at time $t$. For a strong trapping potential, the time-independent component can be expanded up to second order about the mean position $\boldsymbol{\mu}$
\begin{align}
V_{\rm land}(\boldsymbol{r}) &\approx V_{\rm land}(\boldsymbol{\mu}) + (\boldsymbol{r}-\boldsymbol{\mu})^{\top}\nabla V_{\rm land}(\boldsymbol{\mu}) \\ 
&\quad +\frac{1}{2}(\boldsymbol{r}-\boldsymbol{\mu})^{\top} \nabla\nabla^{\top} V_{\rm land}(\boldsymbol{\mu}) (\boldsymbol{r}-\boldsymbol{\mu}) \nonumber \ ,
\end{align}
with $\nabla\nabla^{\top}$ the Hessian matrix. Under these assumptions, the probability distribution can be approximated as Gaussian, $p(\boldsymbol{r},t) \approx \mathcal{N}(\boldsymbol{\mu}_{t},\Sigma_{t})$, with $\boldsymbol{\mu}_{t}$ the average position vector and $\Sigma_{t}$ the covariance matrix at time $t$.

The average entropy produced $\Delta S_{\rm prod} \equiv \Delta S-\beta Q$ (for dimensionless system entropy $S$ and heat $Q$ into the system) in driving from initial probability distribution $p(\boldsymbol{r},0)$ to final probability distribution $p(\boldsymbol{r},\Delta t)$ is~\cite{nakazato2021}
\begin{align}
&\Delta S_{\rm prod} = \int_{0}^{\Delta t} \md t\left[\frac{\md S}{\md t} - \beta \dot{Q} \right] \nonumber\\ 
&= -\int_{0}^{\Delta t} \md t\int \md \boldsymbol{r} \, \frac{\partial p(\boldsymbol{r},t)}{\partial t}\left[\ln p(\boldsymbol{r},t) + \beta V_{\rm tot}(\boldsymbol{r},t) \right] \nonumber\\
&= -\int_{0}^{\Delta t} \md t\int \md \boldsymbol{r}~p(\boldsymbol{r},t)\boldsymbol{v}(\boldsymbol{r},t)\cdot\nabla\left[\beta V_{\rm tot}(\boldsymbol{r},t)+\ln p(\boldsymbol{r},t) \right] \nonumber\\
&= \frac{1}{D} \int_{0}^{\Delta t} \md t \ \langle \boldsymbol{v}(\boldsymbol{r},t)\cdot \boldsymbol{v}(\boldsymbol{r},t)\rangle \ .
\end{align}
Angle brackets $\langle\cdots \rangle$ denote an average over $p(\boldsymbol{r},t)$ and the rate of change of heat is $\dot{Q} \equiv \int \md \boldsymbol{r} V(\boldsymbol{r},t)\partial p(\boldsymbol{r},t)/\partial t$. The second line follows from the standard definitions of entropy and heat,\cite{Seifert2012} the third line from inserting \eqref{Fokker Planck} and integrating by parts over $\boldsymbol{r}$ assuming the probability vanishes at infinity, and the last line from inserting~\eqref{velocity}. From the first law of thermodynamics and the definition of entropy production, the average work is
\begin{align} 
    W = \Delta F_{\rm neq} + \frac{1}{\beta}\Delta S_{\rm prod} \ ,
    \label{Work}
\end{align}
where $\Delta F_{\rm neq} \equiv F_{\rm neq}(\Delta t) - F_{\rm neq}(0)$ is the change in nonequilibrium free energy $F_{\rm neq}(t) \equiv \langle V_{\rm tot}(\boldsymbol{r},t) \rangle - \beta^{-1}\langle \ln p(\boldsymbol{r},t)\rangle $ between the initial and final distributions.

The entropy production is bounded by the squared $L_{2}$-Wasserstein distance between initial and final probability distributions~\cite{nakazato2021}, which for Gaussian distributions is\cite{Olkin1982,abiuso2022}
\begin{align}
    \label{Optimal Entropy}
	\Delta S_{\rm prod} &\geq \\
	&\frac{1}{D\Delta t}\bigg\{ \Delta\boldsymbol{\mu}^2 + {\rm Tr}\left[\Sigma_{0} + \Sigma_{\Delta t} - 2(\Sigma_{\Delta t}^{\frac{1}{2}}\Sigma_{0}\Sigma_{\Delta t}^{\frac{1}{2}})^{\frac{1}{2}}\right]\bigg\}\nonumber \ ,
\end{align}
with subscripts $0$, $t$, and $\Delta t$ respectively denoting the initial, time-dependent, and final values of the corresponding variable. Equality is achieved and the entropy production is minimized when following the optimal-transport map between the initial and final distributions, which for Gaussian distributions is completely specified by the mean and covariance:
\begin{subequations}
\begin{align}
\label{optimal mean}
\boldsymbol{\mu}_{t} &= \boldsymbol{\mu}_{0}+\frac{\Delta \boldsymbol{\mu}}{\Delta t}t \\
\Sigma_{t} &= \left[\left(1-\frac{t}{\Delta t}\right)I +\frac{t}{\Delta t}C \right]\Sigma_{0}\left[\left(1-\frac{t}{\Delta t}\right)I+\frac{t}{\Delta t}C\right] \ .
\label{optimal variance}
\end{align}
\end{subequations}
Here $I$ is the identity matrix, $C \equiv \Sigma_{\Delta t}^{\frac{1}{2}}(\Sigma_{\Delta t}^{\frac{1}{2}}\Sigma_{0}\Sigma_{\Delta t}^{\frac{1}{2}})^{-\frac{1}{2}}\Sigma_{\Delta t}^{\frac{1}{2}}$ reduces in 1D to the ratio of final and initial standard deviations, and $\Delta \boldsymbol{\mu} \equiv \boldsymbol{\mu}_{\Delta t} - \boldsymbol{\mu}_{0}$ is the total change in mean position. If the covariance matrix is diagonal, then \eqref{optimal variance} simplifies to
\begin{align}
    \Sigma_{t}^{\frac{1}{2}} &= \Sigma_{0}^{\frac{1}{2}}+\frac{\Delta \Sigma^{\frac{1}{2}}}{\Delta t} t \ ,
    \label{optimal variance diagonal}
\end{align}
with $\Delta \Sigma^{\frac{1}{2}} \equiv \Sigma^{\frac{1}{2}}_{\Delta t}-\Sigma^{\frac{1}{2}}_{0}$. Thus the standard deviation in each coordinate is linearly driven between the endpoints.

Solving the dynamical equation of motion~\eqref{Fokker Planck} for the time-dependent mean and covariance 
\begin{subequations}
\begin{align}
    \frac{1}{\beta D}\frac{\md \mu_{t}}{\md t} = &K_t (\boldsymbol{\lambda}_t-\boldsymbol{\mu}_t)-\nabla V_{\rm land}(\boldsymbol{\mu}_t) \\ 
    \frac{1}{\beta D}\frac{\md \Sigma_t}{\md t} = &2\beta^{-1}-\left[K_t+\nabla\nabla^{\top} V_{\rm land}(\boldsymbol{\mu}_t)\right]\Sigma_{t} \nonumber\\& - \Sigma_t \left[K_t+\nabla\nabla^{\top} V_{\rm land}(\boldsymbol{\mu}_t)\right]
\end{align}
\end{subequations}
and substituting their optima \eqref{optimal mean} and \eqref{optimal variance}, the trap center and stiffness must respectively satisfy (for a detailed derivation in the absence of an energy landscape see~[\onlinecite{abiuso2022}])
\begin{subequations}
\label{eq:Optima}
\begin{align}
\boldsymbol{\lambda}_t =& \boldsymbol{\mu}_{t} + K_{t}^{-1}\left[\frac{\Delta \boldsymbol{\mu}}{\beta D \Delta t}+\nabla V_{\rm land}(\boldsymbol{\mu}_t)\right] \ ,
\label{optimal center} \\
K_{t}=& \frac{1}{\beta}\Sigma_{t}^{-1} - \frac{1}{\beta D}\int_{0}^{\infty}\md\nu \  e^{-\nu\Sigma_t}\frac{\md \Sigma_t}{\md t}e^{-\nu\Sigma_t} \nonumber \\ 
&- \nabla\nabla^{\top} V_{\rm land}(\boldsymbol{\mu}_t) \ ,
\label{optimal stiffness general}
\end{align}
\end{subequations}
where $\mu_t$ is given by \eqref{optimal mean}, and $\Sigma_t$ by \eqref{optimal variance}. If $\Sigma$ is diagonal, then $\Sigma_t$ is given by \eqref{optimal variance diagonal}, the integral in~\eqref{optimal stiffness general} can be evaluated, and the trap stiffness obeys
\begin{align}
    K_{t} &=\nabla\nabla^{\top} V_{\rm land}(\boldsymbol{\mu}_t) + \left(\frac{1}{\beta}I- \frac{\Delta \Sigma^{\frac{1}{2}} }{2\beta D\Delta t}\right)\Sigma_{t}^{-1} . 
\end{align}

These explicit protocol equations~\eqref{eq:Optima} are considerably easier to compute compared to previous methods for determining minimum-dissipation protocols for CFDs, which require solving differential equations or inverting the Fokker-Planck equation~\cite{Aurell2011,Proesmans2020}. By constraining the final covariance matrix after one period (during which the mean completes one rotation) to equal the initial, we achieve periodic driving: the first two moments are periodic in time. Therefore, to minimize dissipation of a periodic system the covariance remains unchanged throughout the protocol~\eqref{optimal variance}. This is achieved when the effective stiffness is constant, i.e.,
\begin{align}
    K_{t}= K_{0}+\nabla \nabla^{\top} V_{\rm land}(\boldsymbol{\mu}_{0}) -\nabla \nabla^{\top} V_{\rm land}(\boldsymbol{\mu}_{t}) \ .
\label{optimal stiffness}
\end{align}
This results in entropy production 
\begin{align}
\Delta S_{\rm prod} =  \frac{(\Delta\boldsymbol{\mu})^2 }{D\Delta t} \ ,
\label{entropy production at constant variance}
\end{align}
that of an overdamped system moving at constant velocity against viscous Stokes drag; i.e., the minimum-dissipation protocol has perfect Stokes efficiency~\cite{wang2002}. 

For a machine transducing free energy $\Delta F_{\rm neq}$ between the initial and final distributions with equal covariance, the efficiency is the ratio of output free energy to input work,
\begin{align}
\eta \equiv \frac{\beta\Delta F_{\rm neq}}{\beta\Delta F_{\rm neq} + \Delta S_{\rm prod}} \ , 
\end{align}
with the minimum-dissipation protocol achieving the upper bound,
\begin{align}
\eta_{\rm max} = \left[1+\frac{(\Delta\boldsymbol{\mu})^2}{\beta D\Delta t\Delta F_{\rm neq}}\right]^{-1} \ .
\label{efficiency}
\end{align}
Since the entropy production is independent of the free-energy change, a system that travels the same distance but transduces more free energy is more efficient.

\emph{Free-energy estimation.}---Free-energy differences between two equilibrium states of a system can be estimated from nonequilibrium work measurements using the Jarzynski equality or the Crooks fluctuation theorem~\cite{Jarzynski1997,Crooks1999,Jarzynski2011}. The Jarzynski estimator of the free-energy difference $\Delta F_{\rm eq} \equiv F_{\rm eq}[\boldsymbol{\lambda}_{\Delta t},K_{\Delta t}] - F_{\rm eq}[\boldsymbol{\lambda}_{0},K_{0}]$ between equilibrium distributions corresponding to constrained initial and final control parameters is
\begin{align}
    \widehat{\Delta F}_{\rm Jar} = -\frac{1}{\beta}\ln\frac{1}{N}\sum_{i} e^{-\beta W^{(i)}} \ ,
\end{align}
with $W^{(i)}$ the $i$th measurement of work from driving the system, and $N$ the number of samples. In general, the statistical error of the free-energy estimate based on Jarzynski's equality increases with dissipation. The connection between statistical error and dissipation is clearest when dissipation is small, where the expected bias and variance are approximately~\cite{Gore2003,Blaber2020Skewed}
\begin{subequations}
\begin{align}
\label{Bias}
    \langle \widehat{\Delta F}_{\rm Jar}\rangle - \Delta F_{\rm eq} &\approx \frac{1}{N}( W  -\Delta F_{\rm eq}) \\
    \left\langle \left(\widehat{\Delta F}_{\rm Jar} - \left\langle \widehat{\Delta F}_{\rm Jar}\right\rangle\right)^2\right\rangle &\approx \frac{2}{\beta N}( W -\Delta F_{\rm eq}) \ .
\label{Variance}
\end{align}
\end{subequations}
If dissipation is small, minimizing work also minimizes the bias and variance of free energies estimated from Jarzynski's equality. Similar connections can be made for other free-energy estimators such as Bennett's acceptance ratio~\cite{Blaber2020Skewed,Bennett1976,Shirts2003}.

The average work~\eqref{Work} for the minimum-dissipation protocol is
\begin{align}
    W  = & \frac{1}{2}{\rm Tr}\left\{K\left[\Sigma+(\boldsymbol{\mu}-\boldsymbol{\lambda})(\boldsymbol{\mu}-\boldsymbol{\lambda})^{\top}\right]\right\}_{0}^{\Delta t}+V_{\rm land}(\boldsymbol{\mu})\big|_0^{\Delta t} \nonumber \\
    &+ \frac{1}{2}{\rm Tr}\left[\nabla\nabla^{\top} V_{\rm land}(\boldsymbol{\mu})\Sigma\right]_{0}^{\Delta t} + \frac{1}{\beta}\Delta S_{\rm prod}^{\rm min} \ ,
    \label{Optimal Work}
\end{align}
with ${\rm Tr}$ the trace and $\Delta S_{\rm prod}^{\rm min}$ the lower bound in \eqref{Optimal Entropy}. To find the protocol that minimizes work for CFCPs we minimize \eqref{Optimal Work} with respect to the final mean $\mu_{\Delta t}$ and covariance $\Sigma_{\Delta t}$, for fixed final trap center $\boldsymbol{\lambda}_{\Delta t}$ and stiffness $K_{\Delta t}$.

For equal initial and final covariance and a flat energy landscape, the final mean is
\begin{align}
\boldsymbol{\mu}_{\Delta t} = \boldsymbol{\mu}_0 + \left(\frac{2K^{-1}}{\beta D\Delta t} + I\right)^{-1}[\boldsymbol{\lambda}_{\Delta t} - \boldsymbol{\mu}_0] \ .
\label{Optimal Final Mean}
\end{align}
In some more general cases (e.g., energy landscapes represented by low order polynomials), \eqref{Optimal Work} can also be minimized analytically, and in general can be solved numerically with relative ease. Performing this minimization is a considerably simpler task than finding the minimum-dissipation protocols based on thermodynamic-geometry frameworks, which typically require calculating metric tensors and solving geodesic equations~\cite{Blaber2022,louwerse2022}. Another benefit of the present method is that it does not rely on either slow~\cite{OptimalPaths} or fast~\cite{Blaber2021} protocol approximations and is valid at any duration provided the trapping potential is sufficiently strong.

\emph{Rotary motor.}---We demonstrate the applicability of our approximation with a model of a rotary motor inspired by ATP synthase. We consider a one-dimensional periodic energy landscape (Fig.~\ref{Protocol}a),
\begin{align}
	V_{\rm land}(x) = \frac{E_{\rm b}}{2}\left(1-\cos
	\frac{2\pi }{\Delta x _{\rm m}}x
	\right) + \frac{\Delta E }{\Delta x_{\rm m}}x \ .
	\label{periodic landscape}
\end{align}
The barrier height is $E_{\rm b}$, the distance from peak to well is $x_{\rm m}$, the distance between wells is $\Delta x_{\rm m} = 2x_{\rm m}$, and the machine transduces energy $\Delta E$ per barrier crossing. We assume a periodic protocol with equal initial and final variances, $\Sigma_{0} = \Sigma_{\Delta t}$, starting with the mean position at the center of a well, $\mu_{0} = 0$, and terminating after three barrier crossings so that $\mu_{\Delta t} = 3\Delta x_{\rm m}$. For the model's periodic energy landscape~\eqref{periodic landscape} and initial and final means, substituting \eqref{optimal mean} into \eqref{optimal center} and \eqref{optimal stiffness} gives the minimum-dissipation protocol
\begin{subequations}
\begin{align}
    \lambda_{t} =& \frac{1}{k_{t}}\left(\frac{3\Delta x_{\rm m}}{\beta D \Delta t}+\frac{\Delta E}{\Delta x_{\rm m}}+\frac{\pi E_{\rm b}}{\Delta x _{\rm m}}\sin\frac{6\pi }{\Delta t}t\right) \nonumber \\
    &+\frac{3\Delta x_{\rm m}}{\Delta t}t\ ,  \\
    k_{t} =& k_{0}+\frac{2\pi^2 E_{\rm b}}{\Delta x_{\rm m}^2} -\frac{2\pi^2 E_{\rm B}}{\Delta x_{\rm m}^2}\cos\frac{6\pi}{\Delta t}t \ .
\end{align}
\end{subequations}

Figure~\ref{Protocol}a shows the designed intermediate-duration protocol for driving the system over three barriers, numerically estimated from Langevin dynamics integrated with the Euler–Maruyama method~\cite{kloeden1992} with sufficiently small time steps and numerous samples such that numerical inaccuracies are negligible. To maintain constant variance, the designed protocol tightens the trap as it crosses the barriers; to linearly drive the mean between the two wells, the trap center initially jumps ahead of the mean position $\mu_0$, remaining ahead throughout the protocol. For a Gaussian distribution, the $9\%$, $25\%$, $75\%$, and $91\%$ quantiles are evenly spaced, consistent with the linear translation of the quantiles between the two wells, shown in Fig.~\ref{Protocol}a for $k_{0} = 2E_{\rm b}/x_{\rm m}^2$. For a weak trap ($k_{0} = E_{\rm b}/(2x_{\rm m}^2)$), the quantiles are not evenly spaced and exhibit significant deviations from linear temporal evolution, implying that the Gaussian approximation is no longer valid. The crossover from strong to weak trap occurs when $k_{0} \sim E_{\rm b}/x_{\rm m}^2$, since a weaker trap ($k_{0} \lesssim E_{\rm b}/x_{\rm m}^2$) is not sufficient to confine the system within a single well and the distribution can become bimodal, resulting in widely separated quantiles as the system crosses the barriers (Fig.~\ref{Protocol}a).

The quadratic approximation is accurate when the initial stiffness is large ($k_{0} \gg E_{\rm b}/x_{\rm m}^2$). When this condition holds, the entropy production and efficiency are well approximated by \eqref{entropy production at constant variance} and \eqref{efficiency} at any protocol duration (Fig.~\ref{Protocol} b/c). Additionally, for a fast protocol whose duration is shorter than the diffusion time between adjacent wells ($\Delta t \ll \tau_{\rm D} \equiv \Delta x_{\rm m}^2/D$), the entropy production and efficiency agree with \eqref{entropy production at constant variance} and \eqref{efficiency} even for a relatively weak initial stiffness. Large forces are required to rapidly drive the system, which can only be achieved by the trap potential (since the energy landscape is not dynamically controlled), resulting in the dominant contribution to the force arising from the trap potential. Therefore, the approximation is valid when either the protocol duration is short ($\Delta t \ll \tau_{\rm D}$) or the initial stiffness is large ($k_{0} \gg E_{\rm b}/x_{\rm m}^2$). 

Despite the quadratic approximation breaking down when both $\Delta t \gtrsim \tau_{\rm D}$ and $k_{0} \lesssim E_{\rm b}/x_{\rm m}^2$, the final position's mean and standard deviation remain within 20\% of their respective targets, relative to the distance between the wells (Fig.~\ref{Protocol} d/e). Even when the approximations break down, the designed protocols successfully drive the system to the final desired distribution.

\emph{Discussion.}---By approximating static energy landscapes as locally quadratic, we have derived minimum-dissipation protocols for quadratic trapping potentials. This approximation does not rely on either slow or fast limits and therefore offers a complementary result to previous work on designing minimum-dissipation protocols in the fast and slow limits~\cite{Blaber2021,OptimalPaths}. Designed protocols based on the present approximation are considerably simpler than previous methods for determining the minimum-dissipation protocols, which require estimating correlation functions and solving geodesic equations. The trap center linearly drives the mean between the specified endpoints~\eqref{optimal center}; if the initial and final covariances are equal, then the stiffness adjusts to maintain constant covariance throughout the protocol~\eqref{optimal stiffness}. 

We demonstrate the applicability of the approximation with a simple model of a driven rotary motor (Fig.~\ref{Protocol}). When either the initial stiffness is large ($k_{0} \gg E_{\rm b}/x_{\rm m}^2$) or the duration is short ($\Delta t \ll \tau_{\rm D}$), the motor achieves the maximum efficiency~\eqref{efficiency}. When the initial stiffness is small ($k_{0} \lesssim E_{\rm b}/x_{\rm m}^2$) and the duration is large ($\Delta t \gtrsim \tau_{\rm D}$) the motor achieves significantly lower efficiency but the designed protocols still drive the system to within 20\% of the target endpoints relative to the inter-well distance.

Our formalism gives insight into the design principles of efficient motors. Achieving maximum efficiency requires full control of the system, which in general would require an infinite number of control parameters; however, full control of Gaussian probability distributions can be achieved with a finite number of parameters. Within our quadratic approximation, for a $d$-dimensional system the number of control parameters required for an arbitrary energy landscape is $d(d+3)/2$: $d$ trap center components (controlling the means) and $d(d+1)/2$ stiffness matrix components (controlling the covariances). 

We emphasize that the intermediate states remaining Gaussian in the optimal-transport process is the result of an optimization over all possible distributions connecting Gaussian end-states and not an imposed constraint on the intermediate distributions. This is in contrast to parametric methods for determining the minimum-dissipation protocols, where the intermediate states are constrained to those accessible by the small number of control parameters.\cite{OptimalPaths,Salamon1983} In general, we expect the minimum-dissipation protocols determined from parametric control to coincide with optimal transport when there are sufficiently many control parameters to access the intermediate distributions of the optimal-transport process; e.g., $d(d+3)/2$ control parameters for a $d$-dimensional system with a flat energy landscape and a quadratic trapping potential. Otherwise, the full control afforded by the optimal-transport process will achieve less dissipation. 

Several of our results are directly applicable to physical systems. Single-control-parameter (typically the trap center) designed pulling protocols for unfolding DNA hairpins can reduce dissipation~\cite{Tafoya2019}. Our recent theoretical study~\cite{Blaber2022} suggests that dissipation in DNA hairpin experiments can be significantly further reduced by adding one additional control parameter (trap stiffness), but further control (beyond trap center and stiffness) would do very little to reduce dissipation. Within the present framework this is easily understood. Control over the trap center drives the system over the energy barrier between the folded and unfolded state, but cannot prevent the increase in variance as it crosses the barrier. By tightening the trap as it traverses the barrier, the system's variance remains constant and the barrier is effectively eliminated. If the trap is reasonably stiff, then the distribution is approximately Gaussian, and two control parameters are sufficient for full control of this one-dimensional system.

The minimum-dissipation protocols described in the \emph{Free-energy estimation} section can be directly applied to improve estimates of free-energy differences. In steered molecular-dynamics simulations, stiff-trap approximations are commonly employed when estimating free-energy differences~\cite{Park2003,Park2004}; therefore, our method is well situated to improve these estimates. More generally, several enhanced-sampling techniques for free-energy estimation add quadratic potentials to smooth potential-energy surfaces~\cite{Miao2015,Wang2021} or trap intermediate states in umbrella sampling~\cite{Park2014}. There could be connections between our minimum-dissipation protocols and the improved performance from smoothing potential-energy surfaces and optimally spacing intermediate states~\cite{Park2014,Shenfeld2009,Minh2019,Pham2011,Pham2012,Wang2019,Mecklenfeld2017}.

A benefit of the present formalism is that it allows specification of the final distribution by its mean and covariance while using a finite number of control parameters. Previous methods that specified the final distribution using optimal-transport theory required full control over the potential, in principle requiring infinite control parameters. General designs for parametric control typically constrain final control-parameter values but do not actually achieve a specific target distribution. Being able to specify the final distribution is particularly useful for modeling periodic motors like ATP synthase. Fixing equal initial and final covariance, we periodically drive the motor with a high degree of precision (Fig.~\ref{Protocol} d/e) and give insight into the maximum efficiency of such driving. 

Finally, the ease of determining multidimensional designed protocols opens up the possibility to explore a host of new systems, from coupled transport motors pulling cargo~\cite{Leighton2022} to steered molecular-dynamics simulations of complex condensed-matter systems.

\section*{Acknowledgments}
This work was supported by an SFU Graduate Deans Entrance Scholarship (SB), an NSERC Discovery Grant and Discovery Accelerator Supplement (DAS), and a Tier-II Canada Research Chair (DAS), and was enabled in part by support provided by WestGrid (\url{westgrid.ca}) and the Digital Research Alliance of Canada (\url{alliancecan.ca}). The authors thank Jannik Ehrich (SFU Physics) for illuminating discussions and relations to optimal-transport theory, and Matthew Leighton (SFU Physics) for incisive feedback on the manuscript.

\end{document}